\documentclass[lettersize,journal]{IEEEtran}
\usepackage{amsmath,amsfonts}
\usepackage{algorithmic}
\usepackage{algorithm}
\usepackage{array}
\usepackage[caption=false,font=normalsize,labelfont=sf,textfont=sf]{subfig}
\usepackage{textcomp}
\usepackage{stfloats}
\usepackage{url}
\usepackage{verbatim}
\usepackage{graphicx}
\usepackage{cite}
\begin{document}

\title{\textbf{Using Virtual Reality as a Simulation Tool for Augmented Reality Virtual Windows: Effects on Cognitive Workload and Task Performance}
}

\author{ Tianyu Liu, Weiping He, Mark Billinghurst
\thanks{}
\thanks{}}

\markboth{}%
{Shell \MakeLowercase{\textit{et al.}}: A Sample Article Using IEEEtran.cls for IEEE Journals}

\IEEEpubid{}

\maketitle

\begin{abstract}
Virtual content in Augmented Reality (AR) applications can be constructed according to the designer's requirements, but real environments, are difficult to be accurate control or completely reproduce. This makes it difficult to prototype AR applications for certain real environments. One way to address this issue is to use Virtual Reality (VR) to simulate an AR system, enabling the design of controlled experiments and conducting usability evaluations. However, the effectiveness of using VR to simulate AR has not been well studied. In this paper, we report on a user study (N=20) conducted to investigate the impact of using an VR simulation of AR on participants' task performance and cognitive workload (CWL). Participants performed several office tasks in an AR scene with virtual monitors and then again in the VR-simulated AR scene. While using the interfaces CWL was measured with Electroencephalography (EEG) data and a subjective questionnaire. Results showed that frequent visual checks on the keyboard resulted in decreased task performance and increased cognitive workload. This study found that using AR centered on virtual monitor can be effectively simulated using VR. However, there is more research that can be done, so we also report on the study limitations and directions for future work.

\end{abstract}

\section{Introduction}
The use of physical monitors has been a long-standing practice in completing daily work tasks \cite{5}. As tasks become more complex and require higher processing capabilities, users often find it necessary to use multiple monitors to manage multiple windows\cite{1}. The use of large and curved displays allows for the simultaneous display of a considerable amount of content, which is advantageous in terms of the wide human field of view\cite{5}. However, this approach requires more space, incurs higher costs, and results in a reduction in portability. For example, most people use laptops with small monitors on transport such as trains or planes and in noisy outdoor environments\cite{7}; students in libraries and classrooms also find it difficult to carry large monitors. Although the breakthrough of flexible display materials enables displays to be folded and rolled\cite{8}, it still cannot solve the aforementioned problems. Augmented reality (AR) can superimpose virtual information into the physical environment. The latest AR head-mounted display (AR HMD) can register virtual information into the surrounding environment. Users can create and destroy virtual monitors at any time according to their needs, regardless of the physical space. These virtual displays can be adjusted in size and position arbitrarily, allowing them to be configured to meet the needs of different productivity tasks\cite{9}. However, the computational recognition, positioning, and tracking technologies that AR devices rely on are subject to certain errors and latency problems that are difficult to resolve. Furthermore, environmental conditions may also influence the AR effect, and there is a paucity of research in this area.
When using the virtual monitor in AR HMD for experimental or usability evaluation researches, if the surrounding environment is a dynamic and time-varying special scene, such as emergency\cite{11} and natural disaster drills\cite{13}, special weather conditions\cite{14}, and large-scale industrial environments\cite{15}, due to its complexity and variability, it becomes challenging for experimental designers to control variables, making it difficult to repeat experiments. Virtual reality (VR) systems can immerse people in customized virtual environments, thus offering numerous advantages over AR systems\cite{1}. Many variables can be controlled in greater detail in VR, making it highly important to investigate the potential of VR to replicate the AR environment. By leveraging VR's ability to create controlled and reproducible virtual environments, researchers can effectively simulate the conditions and scenes required for their AR studies, overcoming the limitations imposed by the variability of real-world environments. This approach not only streamlines the experimental design process but also enables researchers to conduct more rigorous and reproducible studies, ultimately advancing our understanding of AR technology and its applications. Although many studies have used this simulation approach, its effectiveness has not been specifically evaluated.

\begin{figure}[!t]
\centering
\includegraphics[width=3.5in]{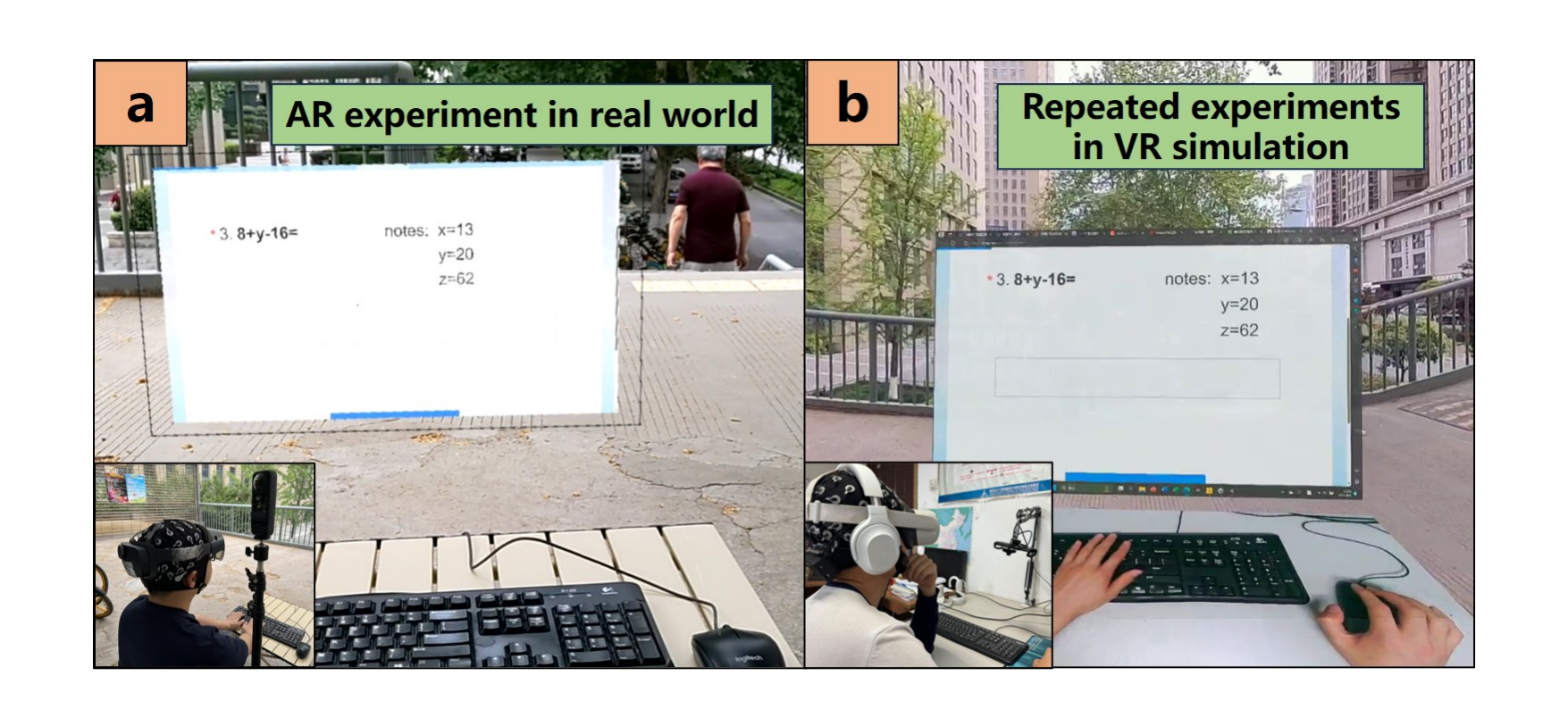}
\caption{Using VR to simulate AR. (a) Participants wearing AR HMD in the real world outdoors (b) Participants use VR indoors to immerse themselves in a simulated world.}
\label{fig_1}
\end{figure}

Our research focuses on the feasibility of utilizing VR to reproduce the environment. The Virtual Reality Reproduction System (VRRS) was developed by us which uses VR to reproduce the background environment and interactive objects in AR office tasks. When a participant employs an AR HMD to construct a virtual monitor for work purposes, the surrounding environment and auditory stimuli are recorded. The VRRS can generate the same virtual monitor, and play the 360° background environment video and the real-time video of the captured keyboard and mouse. A series of office tasks were designed based on AR HMD and VRRS, and a user study was conducted to compare the impact of the two conditions on participants’ cognitive workload (CWL) and task performance. Considering that the use of virtual screen office is a common function of AR HMD, these tasks are completed based on the virtual screen. We employed Electroencephalography (EEG) to measure brain activity and data pertaining to automation, including task duration and results, was collected. To date, no research has been identified that compares the use of VR systems to reproduce the interactive content and environment when using AR office systems. The purpose of this paper is to verify the feasibility of the widely used method of simulating AR using VR. Fig.\ref{fig_1} shows how we simulate real-world scenarios.  

So, our research makes the following contributions:

(1) We have evaluated the similarities and differences in cognitive load between users using the VRSS and those using AR HMD.

 (2) We discussed the differences with respect to usability and other aspects.
 
(3) We provided some recommendations for using a VR system to reproduce the environment of an AR system.

This article is organized as follows: Section 2 presents related work on the utilization of VR to reproduce the AR task and environment. Section 3 describes the building of the VRSS and our task setup and implementation process. Section 4 describes a user study to evaluate these AR instruction interfaces, while Section 5 presents the results. Section 6 discusses our results. Finally, Section 7 concludes and summarizes this article.

\section{Related work}
We aim to leverage the insights from two main bodies of research: \textit{Effects of real-world environment in AR task} and \textit{Simulating AR with VR}. Therefore, in this section, we first seek to understand the impact of the environment on AR. Then, we describe how previous systems simulated AR with VR.

\subsection{Effects of real-world environment in AR task }
The use of AR HMD has been explored as an alternative to physical monitor\cite{5}. Pavanatto et al. found that virtual monitors constructed by AR HMDs could be used for real-world productivity work in a short period with guaranteed task accuracy\cite{1}. A long-term goal of AR researchers is to make it possible for the real and virtual worlds to blend together seamlessly. However, humans are actively engaged in all the different modalities perceived at any moment, and in AR applications, saturation may be reached due to humans' limited cognitive abilities\cite{5}. Therefore, it is necessary for researchers to divide AR scenes into directly interactive components and real background environments\cite{6,12,16,30} to avoid confusion. The directly interactive components include physical objects and virtual objects, while the background environment will interfere with the participants.

Nenna et al. evaluated the cost of dual-tasking on cognitive and motor performance in AR task\cite{17}, Participants were tested outdoors in a university campus back square. They found that the attentional load induced by multitasking has significant implications for navigating the dynamic real world. Therefore, it is essential to address this issue in a variety of daily-living contexts. Attention to the background environment when using AR systems can cause users to switch context and refocus their eyes, Gabbard et al. developed a structured visual search task that required participants to switch both context and focal distance between real and AR virtual text\cite{18}. They proposed that performance was enhanced when all information was available in the real world, as opposed to when users had to integrate information between the real and virtual spaces. In their discussion of the educational implications of AR learning environments, Geana et al. note that in a familiar environment, learners can engage with virtual components at their own pace, thus enhancing the overall learning experience\cite{19}. The findings of these studies collectively indicate that those responsible for the design of AR applications or experiments must give full consideration to the impact of the real environment on the user. 

Ideally, an AR monitor should work in all environments without the need to prepare rooms ahead of time\cite{21}. It is necessary to conduct research into the potential of AR to function in unstructured real-world environments. Leder et al. developed a system for placing and integrating literature 3D models into real environments\cite{22}. They indicated that outdoor AR applications can be employed in different use cases, with the potential to enhance the comprehension of architectural visualizations. Furthermore, several requirements were defined in terms of the nature of outdoor usage, including weather and lighting robustness and mobile usage. Gabbard et al. indicated that outdoor background textures and natural lighting influence user performance in outdoor AR\cite{31}.

Furthermore, the integration of multi-modal information in real environments can have a multitude of effects on users engaged in AR tasks. Zhou et al. developed a system to examine the impacts of 3-D sound on improving depth perception and shortening task completion time\cite{23}. They found that 3-D sound is an effective complement to visual AR environments, which can enhance task performance and facilitate collaboration between users. 3-D sound also enables a more realistic environment and a more immersive experience of being inside the AR environment, through both visual and auditory means. Feng et al. developed an AR-Assisted Cooperative Assembly System\cite{24}. They used noise to simulate the real riveting environment. 

\subsection{Simulating AR with VR}
VR offers users the opportunity to engage in immersive simulations within defined virtual environments, which can confer numerous advantages to experimental design and training. Its applications are diverse, spanning healthcare\cite{24}, education\cite{26}, and industry\cite{27}.  The utilization of VR to simulate the interactive components and background environment of an AR system enables researchers to regulate the conditions for comparative experiments, or even to test systems or scenes that do not exist. There are generally two ways to use VR to simulate AR: building a virtual environment and replaying the real environment.

Lee et al. replicated a well-known AR study by Ellis et al. using their VR simulator\cite{28}. Svensson developed a system that simulates location-based AR storytelling in a VR environment\cite{29}. The system was used for testing local news distribution with AR devices. Furthermore, it can be employed as a more general tool for testing the potential applications of augmented reality in other contexts. Tran et al. developed a system that enables participants to interact in a controlled VR environment to simulate the experiences of wearing urban AR devices\cite{30}. They conducted two simulations: pedestrian navigation and autonomous mobility. It was demonstrated that the utilization of immersive VR to simulate the functionality of wearable AR applications within an urban context is an effective approach. Kang et al. developed a fire drill training system which can improve training effectiveness by transforming arbitrary real spaces into real-time, realistic virtual fire situations, and by interacting with tangible training props\cite{32}. Lenz et al. studied asynchronous task interaction in MR systems\cite{42}. They also used VR simulation to control experimental conditions in their research. Creating virtual environments has been widely used, but researchers have mostly used this simulation method for experiments or evaluations, and have not studied the usability of this method.

Sylaiou et al. created a virtual museum with VR and AR elements, they found a positive correlation between enjoyment and both VR and AR object presence\cite{33}. See et al. who the investigated user experience of a boat builders' AR and VR showcase pillar in Pangkor. Their result rated the 360° VR experience as more natural and useful than AR's static image and text\cite{34}. Verhulst et al. compared the user experience of AR and VR versions of an immersive cultural experience, the scenes in VR are captured in reality and kept very consistent with AR. They found that VR outperformed AR on enjoyment and most presence items\cite{35}. VR simulation is not only used for experimental design and training but also can be used in remote collaboration. Teo et al. developed a 360° video and 3D reconstruction mixed system for remote collaboration\cite{37}.  Piumsomboon et al. developed a system combining AR, VR, and natural communication cues to create new types of collaboration\cite{38}. Wang et al. developed a gesture-based remote collaborative platform. In these remote collaboration systems, the AR user's real environment is simulated for the VR user in different ways, making them feel like they are sharing the same space. Replaying real environments has also been used by many participants, but the effects on participants have not been studied individually.

Our work differs from the above in focusing on evaluating the effectiveness of VR for Simulating AR Environments. We reproduce the environment by replaying the background environment and enhancing interactive content. Using a virtual screen is considered a common feature of AR HMDs, so we focused on evaluating this scenario to explore whether it can provide convenience for experiment designers. In addition, to obtain more objective data, we also checked the participants' EEG during the process.

\section{Virtual Reality Reproduction System setup}
Using VR to replicate scenes in which users perform AR tasks serves as a guide for the design of controlled experiments and provides repeatable conditions for experiments that have been researched extensively. In today’s workplace, the physical QWERTY keyboard and mouse configuration continues to prevail\cite{69}, while AR HMDs present a promising avenue for visualization\cite{39}. We designed the Virtual Reality Reproduction System (VRRS) where the participant’s goal was to complete a range of office tasks including typing, CAD drawing, etc. An AR HMD was used to create a virtual screen and complete tasks in high-interference environment. At the same time, the background environment was recorded. In VRSS, the background environment would be replayed, and the screen, keyboard, and mouse with which they interact directly were also recreated. In addition, in order to evaluate the impact of the device on the participants separately, we also set up a low-interference environment, which will also be replicated by VRSS for another experiment.
  \begin{figure}[!t]
\centering
\includegraphics[width=3.5in]{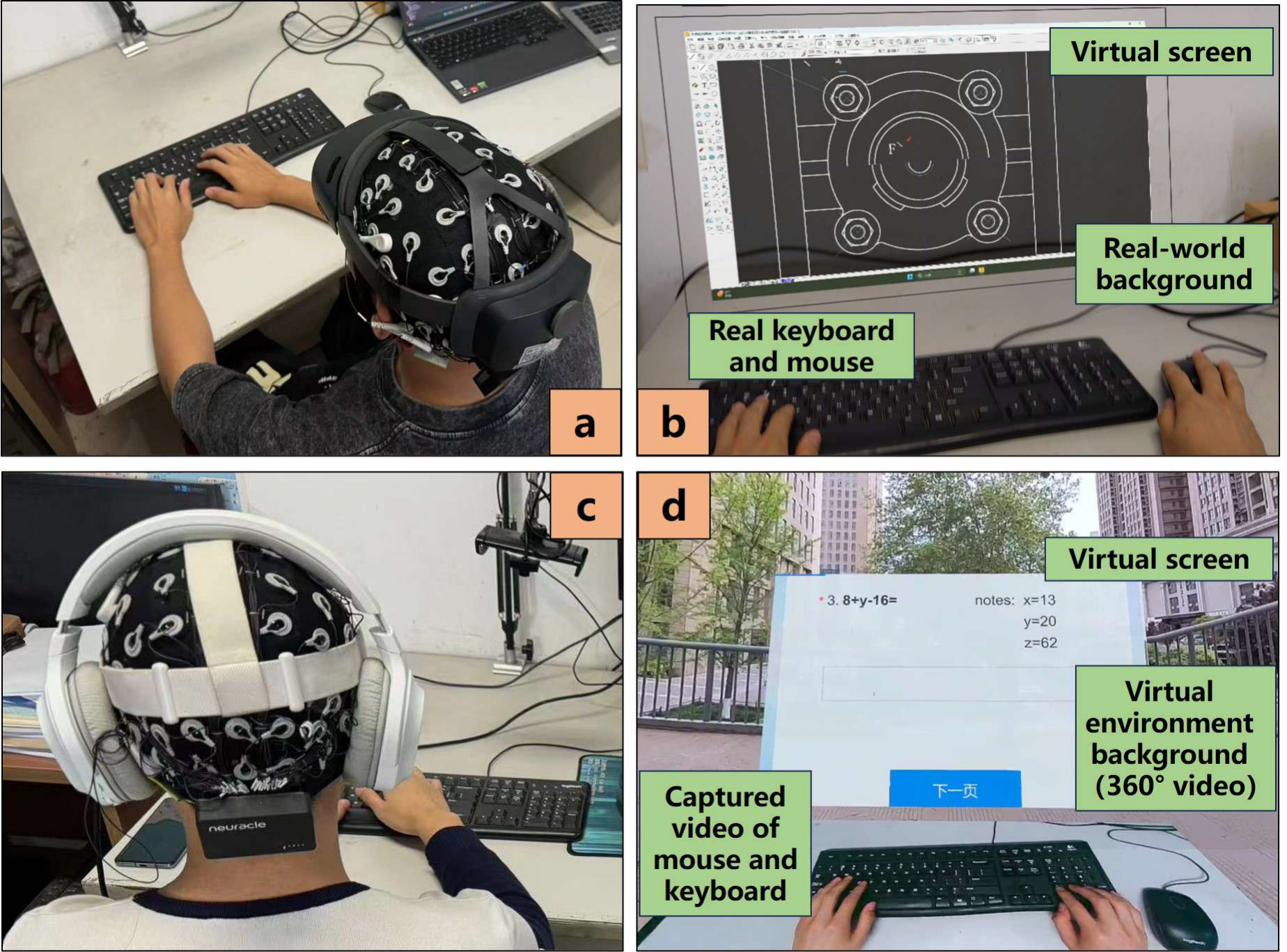}
\caption{AR and VRSS conditions. (a) Participant using AR HMD. (b) Participant’ s perspective under AR conditions. (c) Participant using VRSS. (d) Participant’ s perspective under VRSS conditions.}
\label{fig_2}
\end{figure}

\subsection{VRSS design}
 Fig.\ref{fig_2}b shows the office scene we set up for AR HMD users: a virtual screen, mouse, and keyboard are provided to the user, along with an independent desk and chair. Additionally, users are ensured not to be disturbed during the entirety of the experiment. The SLAM function of AR HMD ensures that the relative position of the virtual screen and the environment remains unchanged, thereby imitating the real physical screen. However, the virtual screen is unable to cover the environment due to its optical characteristics. During the entire process of the task, the background environmental images and sounds were simultaneously captured by the 360° camera and stereo microphone situated in close proximity to the user's head.

 Fig.\ref{fig_2}d shows an overview of the VRSS setup. The user is immersed in the VR scene. The dimensions, position, and transparency of the virtual screen are set to the same as in AR condition. A window plays the video of the mouse and keyboard being captured by the camera, which had proven successful in VR\cite{42}. The background environment is shown in VR, with stereo sound played through headphones. Although videos captured by 360° cameras lack depth information, portability and economy can be greatly satisfied by this easily available commodity hardware\cite{40}. The processing requirements for 360° video are typically less demanding than those for point clouds or model reconstructions, as the video data is already in a visual format and does not require the application of complex reconstruction algorithms \cite{41}.

\subsection{hardware and software}
\begin{figure}[!t]
\centering
\includegraphics[width=3.5in]{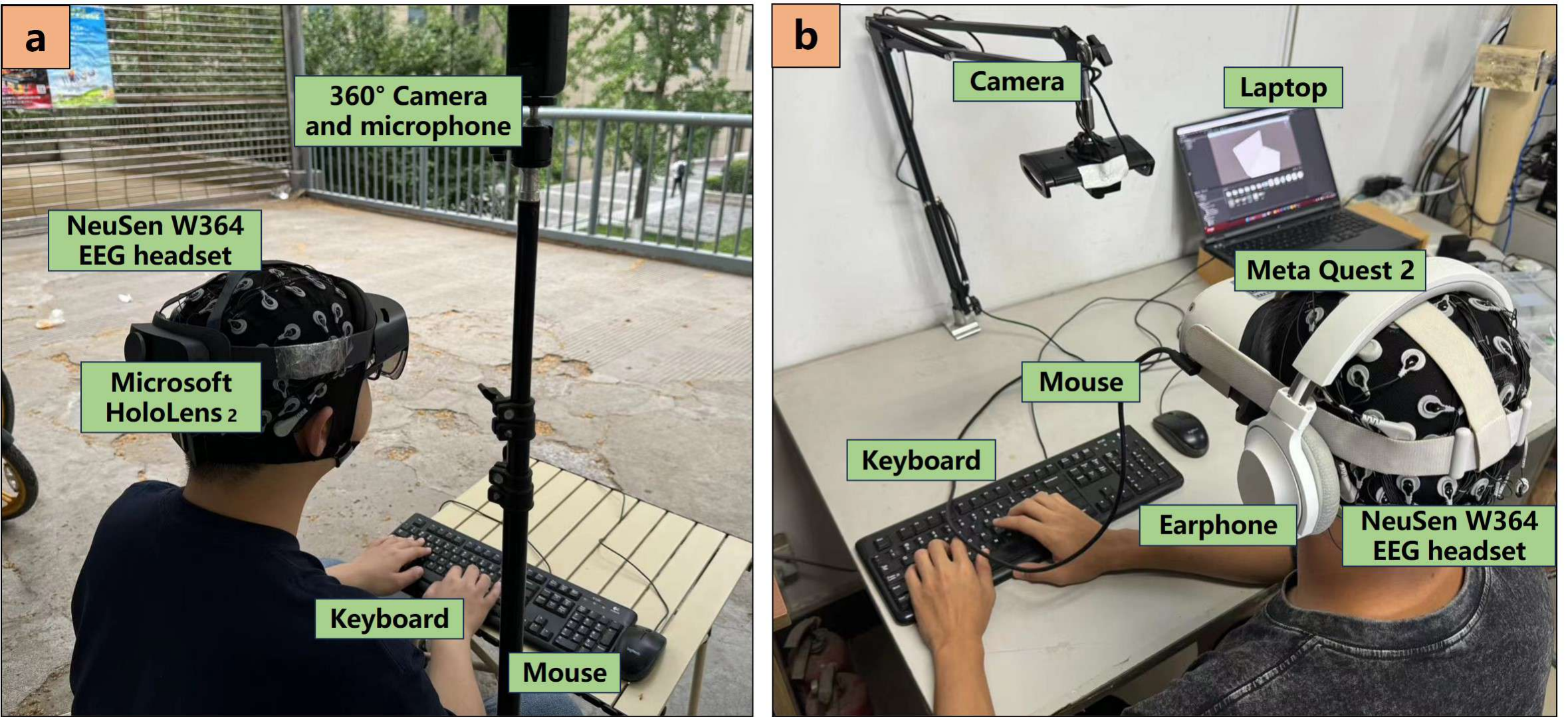}
\caption{Overview of hardware. (a) Hardware setup when using AR HMD. (b) Hardware setup when using VRSS.}
\label{fig_3}
\end{figure}

The hardware is shown in Fig. \ref{fig_3}. When participants performed AR tasks, the equipment used was Microsoft HoloLens 2 \cite{47}optical see-through AR HMD with a 52◦ diagonal FOV (43° horizontal and 29° vertical). The input devices employed were a Logitech K120 USB Standard Computer Keyboard\cite{48}, a full-size keyboard with an integrated number pad, and a Logitech M100 USB corded mouse\cite{49}. The VRSS comprised a Meta Quest 2 VR headset \cite{50}with an LCD display, with a resolution of 1832×1920 per eye, 90 Hz refresh rate, and a USB camera with 1920×1080 image resolution. The VR headset was connected to a laptop with the following specification via a USB-C cable: AMD Ryzen 7 5800H Core, NVIDIA GeForce RTX 3060 Laptop GPU, 16GB DDR4 Quad Channel DDR4 memory, 512GB SSD. The operating system was Windows 11 x64. The virtual screen was created with Mirage APP \cite{51}in HoloLens 2. The VRSS was implemented in the Unity3D engine 2019.4.40 \cite{52}with C\#. To record 360° video, an Insta 360X3 camera\cite{53}was used with a video resolution of 5760x2880@30. DJI Mic\cite{54} was used to record stereo audio and then played back using a headphone. To measure EEG data, a 64-channel NeuSen W364\cite{55} wireless EEG equipment, which is an international 10-20 electrode position, was used.

\section{User Study}
\subsection{research question}
Although many researchers have proposed and applied VR simulations to design AR experiments\cite{1,28,29,30}. However, there has been no research into the effectiveness of VR rendering for AR scenes with high levels of environmental dynamic interference. The VRSS we proposed aims to quickly and effectively reproduce dynamically disturbed AR office scenes to help conduct controlled experiments. However, it is not clear whether the proposed method is feasible for designing user experiments. Therefore, the research question is:
\textit{RQ: Can the method of  VR simulation scenes effectively reproduce AR office work?}

   \begin{figure}[!t]
\centering
\includegraphics[width=3.5in]{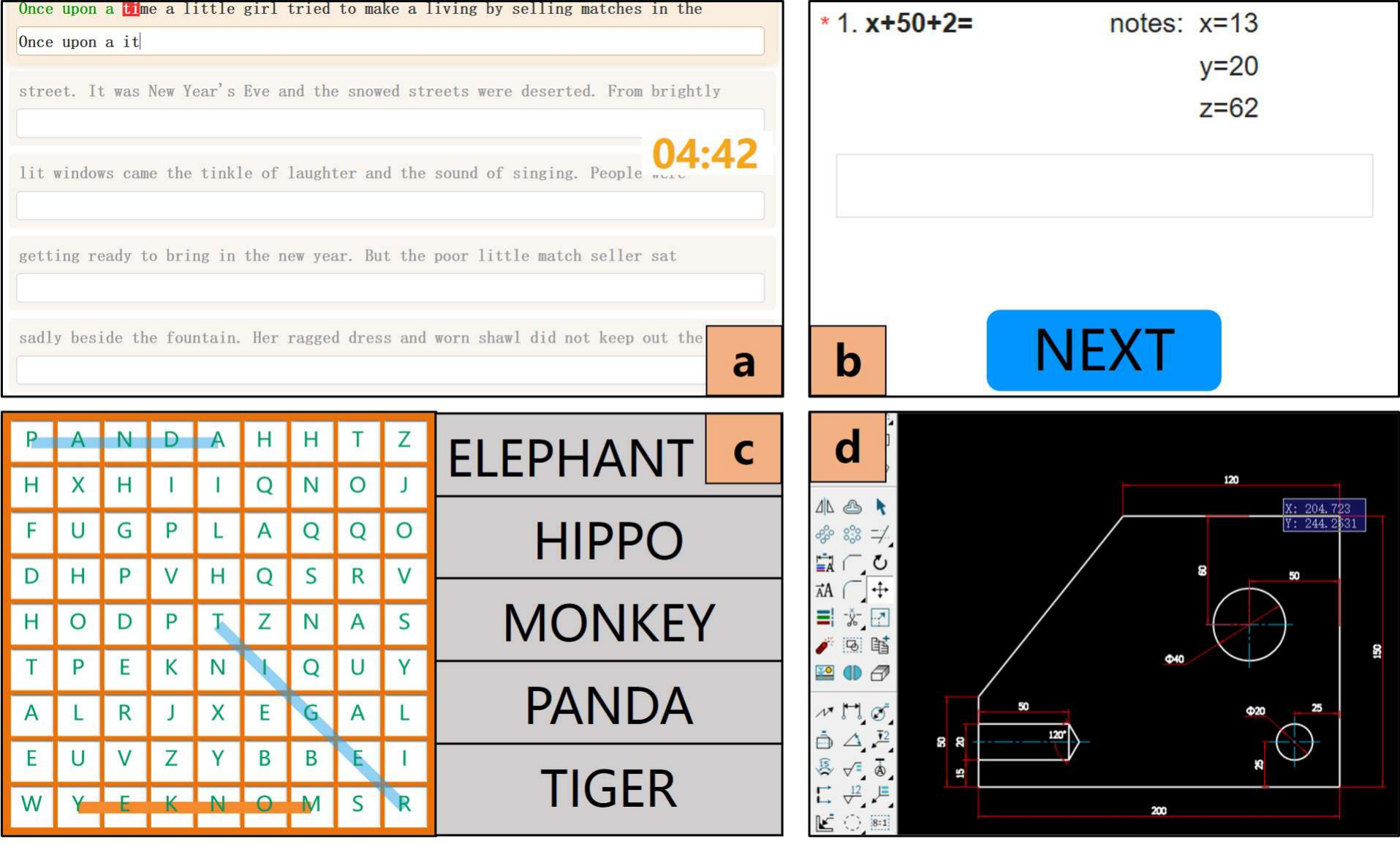}
\caption{Interfaces for the tasks. (a) Correct typing will be marked in green, and incorrect typing will be marked in red. (b) The variable will be displayed on the right and used to calculate the expression on the left. (c) A correct connection will turn blue, and an unfinished connection will remain orange. (d) Graphic features include lines, circles, holes, etc.}
\label{fig_4}
\end{figure}
\subsection{study task }
Previous research on office tasks \cite{42,43,44,45}in VR/AR inspired our study task setup. We designed four tasks to be completed under two conditions: AR and VRSS.

Typing task (Fig. \ref{fig_4} a). Participants were asked to complete the typing task continuously within a 5-minute time limit, and a countdown was displayed. In the input box, if the word entered by the participant is wrong, it will be marked in red to prompt the participant to correct it. 

Search task (Fig. \ref{fig_4} b). Participants were asked to find 5 given words on a 9×9 table. There is one letter in each cell and each word is randomly arranged horizontally, vertically, and tilted ±45°. Letters are not reused after a successful match. After completing the 5-word search task, the positions of the words and letters were updated, and participants had to complete the task a total of 3 times.

Math task (Fig. \ref{fig_4} c). The participant had to calculate the value of 20 mathematical expressions of various difficulty levels without using paper and pencil. Some expressions contain variables, and the highest calculation amount is three two-digit arithmetic operations. Only one question will appear on the screen at a time, and participants will not be told whether their answer is correct or incorrect after they have submitted it, nor will they be able to go back and make changes until all the calculated questions have been completed.

Drawing task (Fig. \ref{fig_4} d). Participants must draw a given 2D drawing using the DaXiong CAD software (trained before the experiment). Copying of case graphics is not allowed. Participants must check whether they are correct after completing the drawing and submit it if they believe it is accurate.

\subsection{participants}
We recruited a total of 20 participants (17 males, 3 females; aged 22-26 years, M = 23.55, SD = 0.973). All participants were majors in Mechanical Engineering and Manufacturing and had basic assembly experience. They had normal or corrected-to-normal vision, with no color blindness. Of the participants, 14 were familiar with AR or VR interfaces, providing a rating of three or higher on a 5-point Likert item (1: novice, 5: expert), and no one never experienced VR/AR with a head-mounted display. This indicated that the participants were familiar with AR or VR, which won’t impact their feedback.

\subsection{procedure }
\begin{figure}[!t]
\centering
\includegraphics[width=2.5in]{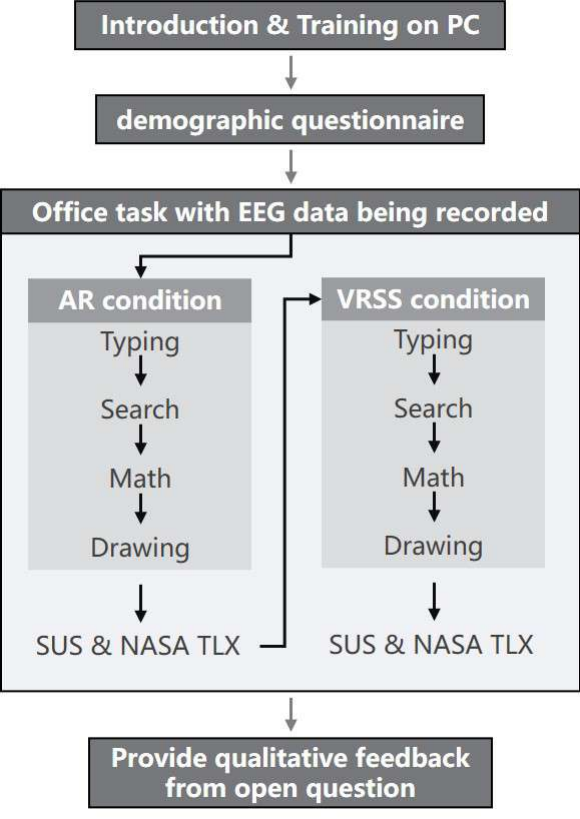}
\caption{The procedure of the user study.}
\label{fig_5}
\end{figure}

With the approval of the Medical and Experimental Animal Ethics Committee of Northwestern Polytechnical University, a within-subjects user study was conducted with the two conditions mentioned above: AR and VRSS. Fig.\ref{fig_5} is an overview of our study procedure. As mentioned above, we set up two scenes in AR and VRSS: an indoor scene and an outdoor scene, each corresponding to the low-interference environment and the high-interference environment, the order of the two environments was balanced. The indoor scene is a quiet office, while the outdoor scene is a pedestrian bridge outside our campus. In the experiment of each participant, we did our best to ensure that the weather and temperature in the outdoor scene were similar, and we concentrated on conducting outdoor experiments between 11:00 and 12:00 on weekdays, as pedestrian and vehicle traffic under the flyover was roughly similar at this time.

Participants were first given an introduction to the experimental procedures and a consent form, and after signing the consent form, they were asked to complete a demographic questionnaire. The experimenter completed the pre-experiment training we set up on the PC, including detailed explanations of the above four tasks. Considering the subject background of the experimenter, we selected DaXiong CAD software\cite{56} and conducted training until the participants were familiar with the basic operations. They were then asked to complete the AR tasks in the office and then go outside to complete the AR tasks when they felt comfortable. After completing the tasks, they were asked to complete a post-experiment questionnaire. Participants then returned inside, completed the tasks when the environment of the two scenes was reproduced in VRSS, and finally completed the post-experiment questionnaire. Throughout the experiment, all questionnaires were completed on a PC, and participants wore an EEG cap during the office tasks and had their EEG data collected. The entire experiment lasted approximately 3 hours, and participants received \$15 worth of gifts as compensation.

\subsection{metrics}
The demographics questionnaire included only a few questions (i.e., name, age, and gender identity). The post-experiment questionnaire included the SUS questionnaire\cite{57} and the standard NASA TLX questionnaire\cite{58}. We recorded typing speed (words per minute) and number of backspaces for the typing task, completion time for the search task, completion time and correct number for the math task, and completion time for the drawing task.

As the demands of tasks performed by humans increase, so does the utilization of brain resources. This utilization of brain resources is called cognitive workload(CWL)\cite{59}. EEG has been employed extensively to elucidate the interrelationship between cognitive workload, attention, and information flow through disparate cognitive systems. A number of previous studies have demonstrated that $\theta$ (4–8Hz) frequencies and $\alpha$ (8–12Hz) frequencies are significantly correlated with cognitive load\cite{60,61}. Specifically, the increase in CWL brings about an increase in theta power and a decrease in alpha power.

We collected EEG data with the NSW headband throughout the experiment. The NSW364 is a 64-channel EEG device. We used PZ, P3, P4, FZ, F3, and F4 electrodes with a sampling frequency of 500Hz. Participants were asked to rest with their eyes closed for 90 seconds before the task.

\section{Results}
We analyzed quantitative data and qualitative data separately: quantitative data included participant performance and EEG results, while qualitative data included results from the SUS and NASA TLX questionnaires.

\subsection{quantitative results}
\subsubsection{participant performance}
\begin{figure*}[!t]
\centering
\includegraphics[width=6in]{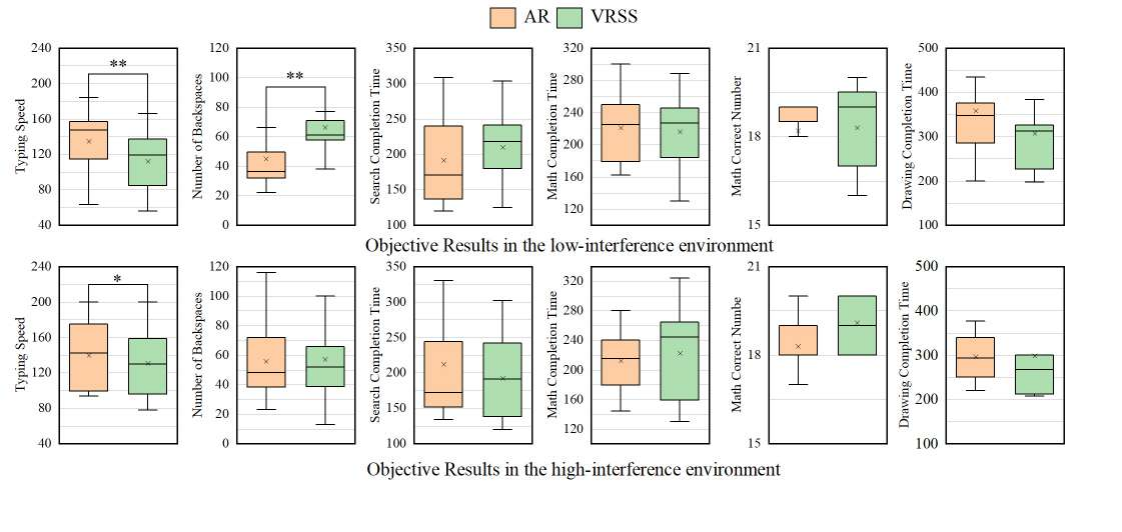}
\caption{The task performance results. $\ast$$\textless$0.05 $\ast$$\ast$ $\textless$0.01}
\label{fig_6}
\end{figure*}

The objective results are shown in Table \ref{tab:table_1}, Table \ref{tab:table_2} and Fig.\ref{fig_6}. The Kolmogorov–Smirnov tests show that all task completion times and typing speeds do not conform to the normal distribution, so they were subjected to Wilcoxon signed-rank tests.

\begin{table*}[!t]
\caption{Mean (standard deviation) of objective results in the low-interference environment.
\label{tab:table_1}}
\centering
\begin{tabular}{lllllll}
\hline
Condition & \begin{tabular}[c]{@{}l@{}}typing speed\\ (words per minute)\end{tabular} & Number of backspaces & \begin{tabular}[c]{@{}l@{}}Search completion\\ time (s)\end{tabular} & \begin{tabular}[c]{@{}l@{}}Math completion\\ time(s)\end{tabular} & Math correct number & \begin{tabular}[c]{@{}l@{}}Drawing completion \\ time(s)\end{tabular} \\ \hline
AR        & 134.65 (34.51)                                                            & 45.05 (23.34)        & 191.20 (63.49)                                                       & 221.00 (44.66)                                                    & 18.20 (2.07)        & 357.90 (117.25)                                                       \\
VRSS      & 112.20 (34.25)                                                            & 66.55 (27.27)        & 210.10 (52.30)                                                       & 216.05 (49.02)                                                    & 18.30 (1.49)        & 307.10 (90.53)                                                        \\ \hline
\end{tabular}
\end{table*}

\begin{table*}[!t]
\caption{Mean (standard deviation) of objective results in the high-interference environment.
\label{tab:table_2}}
\centering
\begin{tabular}{lllllll}
\hline
Condition & \begin{tabular}[c]{@{}l@{}}typing speed\\ (words per minute)\end{tabular} & Number of backspaces & \begin{tabular}[c]{@{}l@{}}Search completion\\ time (s)\end{tabular} & \begin{tabular}[c]{@{}l@{}}Math completion\\ time(s)\end{tabular} & Math correct number & \begin{tabular}[c]{@{}l@{}}Drawing completion \\ time(s)\end{tabular} \\ \hline
AR        & 139.65 (36.21)& 55.80 (25.39)& 211.90 (95.26)& 212.55 (40.60)& 18.30 (1.08)& 296.15 (53.99)\\
VRSS      & 131.00 (38.20)& 57.15 (26.25)& 191.90 (60.04)& 222.55 (59.11)& 19.10 (0.85)& 297.75 (107.65)\\ \hline
\end{tabular}
\end{table*}

The Wilcoxon signed-rank tests showed that the typing speed under AR condition of low-interference environment ($Z$= -3.059, $p$$\textless$0.01) was significantly higher than VRSS condition, and the number of backspaces ($Z$= -2.354, $p$$\textless$0.01) was also significantly less. Furthermore, the typing speed under AR conditions of high-interference environment ($Z$= -2.118, $p$$\textless$0.05) was also significantly higher than that under VRSS conditions. The remaining task performance results showed no significant differences between the AR condition and the VRSS condition. It can be concluded that there was no significant difference in the performance of the aforementioned office tasks under VRSS and AR conditions, with the exception of the typing task.

\subsubsection{EEG data}
The EEGLAB in MATLAB was employed to process the EEG data with the objective of removing noise components such as blinking, clenching the jaw, and turning the head. Subsequently, the $\alpha$ frequency (8-12Hz) and $\theta$ (4-8Hz) frequency were analyzed separately.  we computed a Hedges’s g statistic for the power spectral density (PSD) between the VRSS condition and the AR condition, which was adjusted for the small sample bias in comparison to Cohen’s d\cite{67}. The EEG data of the 16 subjects was finally retained, and the data of the remaining subjects was attributed to invalid data due to electrode connection. The data of each individual task were analyzed separately, and the 90s of rest before the experiment were also removed. The results are shown in Table \ref{tab:table_3} and Table \ref{tab:table_4}. 
We used a correlation value of  r=0.50, which is considered relatively conservative\cite{68}. In the low-interference environment, the CWL of the typing task under the VRSS condition was higher than that under the AR condition, as evidenced by both the $\alpha$  frequency and $\theta$ frequency. In the high-interference environment, the CWL of the typing task under the VRSS condition was much higher than that under the AR condition, as evidenced by the $\alpha$ frequency. In the remaining tasks, the CWL under the VRSS condition and the AR condition were considered to be no different.

\begin{table}[!t]
\caption{Hedges’s g (standard error) of EEG results in the low-interference environment.
\label{tab:table_3}}
\centering
\begin{tabular}{lllll}
\hline
Frequency& Typing& Search& Math& Drawing\\ \hline
$\theta$(4-8Hz)& 0.62 (11.84)& 0.12(13.62)& 0.190(12.72)& -0.05(14.31)\\
$\alpha$(8-12Hz)& -0.65(9.65)& -0.06(13.88)& -0.48(10.75)& -0.41(9.42)\\ \hline
\end{tabular}
\end{table}
\begin{table}[!t]
\caption{Hedges’s g (standard error) of EEG results in the high-interference environment.
\label{tab:table_4}}
\centering
\begin{tabular}{lllll}
\hline
Frequency& Typing& Search& Math& Drawing\\ \hline
$\theta$(4-8Hz)& 0.63(13.32)& 0.04(13.48)& 0.10(14.09)& 0.34(11.28)\\
$\alpha$(8-12Hz)& -0.32(11.02)& -0.29(12.40)& -0.12(11.94)& -0.11(11.55)\\ \hline
\end{tabular}
\end{table}

\subsection{qualitative results }
The subjective data comprised the SUS and the NASA TLX survey. Similarly, these discrete variables were also subjected to a Wilcoxon signed-rank test in order to verify their significant differences.

\subsubsection{SUS}
\begin{figure}[!t]
\centering
\includegraphics[width=3.5in]{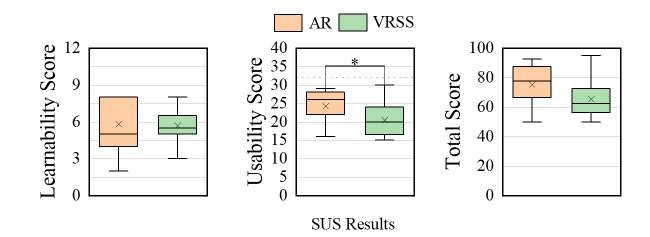}
\caption{Average results for the SUS questionnaires, and the dotted lines show the full scores of the scales. $\ast$$\textless$ 0.05 }
\label{fig_7}
\end{figure}

The results are shown in Table \ref{tab:table_5} and Fig\ref{fig_7}. The SUS usability score ($Z$=-2.380, $p$$\textless$0.05) of AR was significantly higher than that of VRSS. There was no significant difference in the SUS learnability score ($Z$=-0.478, $p$=0.632) and the SUS total score ($Z$=-1.798, $p$=0.072) between VRSS and AR. Sauro et al. calculated an average of 68 from many SUS questionnaires\cite{66}. In terms of system usability, VRSS is below average, while AR is above average. The results indicated that using VRSS to simulate AR will reduce the perceived ease-of-use of AR office.
\begin{table}[!t]
\caption{Mean (standard deviation) of SUS results.
\label{tab:table_5}}
\centering
\begin{tabular}{llll}
\hline
Condition& SUS learnability& SUS usability& SUS total\\ \hline
AR& 72.50(25.84)& 75.94(13.22)& 75.25(13.98)\\
VRSS& 71.25(18.63)& 64.06(14.36)& 65.50(12.42)\\ \hline
\end{tabular}
\end{table}

\subsubsection{NASA TLX}
   \begin{figure*}[!t]
\centering
\includegraphics[width=6in]{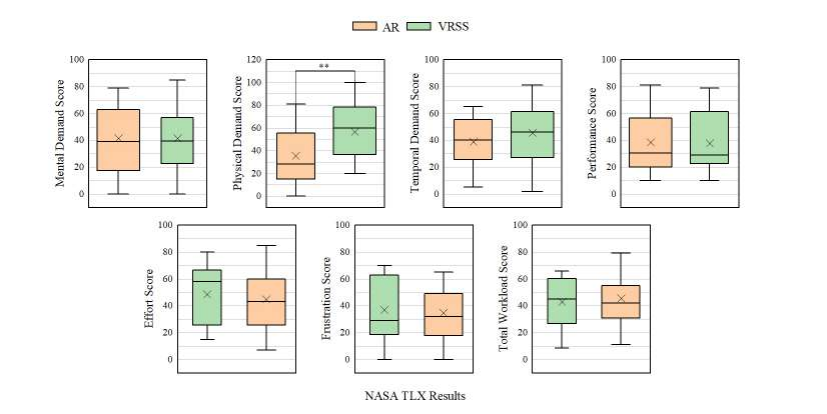}
\caption{Average results for the NASA TLX questionnaires.  $\ast$$\ast$$\textless$0.01}
\label{fig_8}
\end{figure*}

\begin{table*}[!t]
\caption{Mean (standard deviation) of the NASA TLX results.
\label{tab:table_6}}
\centering
\begin{tabular}{llllllll}
\hline
Condition& Mental demand& Physical 
demand& Temporal 
demand& Performance & Effort& Frustration&Total\\ \hline
AR& 41.60 (26.03)& 35.40 (26.29)& 38.75 (16.87)& 38.60 (23.88)& 48.40 (22.58)& 36.85 (23.37)&42.80 (18.76)\\
VRSS& 41.50 (27.03)& 58.10 (24.32)& 45.40 (23.54)& 37.85 (23.12)& 44.95 (23.02)& 34.60 (18.96)&45.59(19.47)\\ \hline
\end{tabular}
\end{table*}

The results are shown in Table \ref{tab:table_6} and Fig\ref{fig_8}. There are significant differences in physical demand ($Z$=-2.959, $p$$\textless$0.005) and no significant differences in mental demand ($Z$=-0.403, $p$=0.687), temporal demand ($Z$=-1.352, $p$=0.177), performance ($Z$=-0.738, $p$=0.461), effort ($Z$=-0.741, $p$=0.459), and frustration ($Z$=-0.305, $p$=0.760), as well as the total workload score ($Z$=-1.419, $p$=0.156), between the two conditions. The results indicated that using VRSS to simulate AR has basically no impact on the perceived workload. 

\section{Discussion}
This section discusses the results of the user research conducted in the previous section, as well as subjective user interviews and observations of user behavior. It also analyzes the limitations of this work and suggests potential avenues for future work.

\subsection{AR vs VR simulation}
This study developed a methodology for simulating AR using VR. In order to evaluate the usability of the system, we focused on the scene of using a virtual screen to perform office tasks, which is considered to be a highly used function in today's AR HMD. The participants were first required to create a virtual screen using an AR HMD and then to complete four office tasks. Subsequently, VR was employed to simulate the background environment and interactive objects (including a virtual screen, keyboard, and mouse), with participants completing the relevant tasks once more. As mentioned above, both objective and subjective statistics were collected to analyze the efficiency, ease-of-use, and CWL. This subsection discusses the results of the user study based on our observations and subjective user interviews.

As for the objective results, the performance of the participants was significantly different only in the typing task. Specifically, the use of VRSS resulted in a decrease in typing speed and an increase in the number of backspaces, particularly in the low-interference environment. Our observations indicated that the typing task differed from other tasks in that it involved a high frequency of interaction with the keyboard. Only participant P14 could achieve touch typing in the typing task (including VRSS condition and AR condition). The remaining participants had to look down at the keyboard to confirm the position of their fingertips after completing the spelling of a few words. The EEG data can fully reflect the CWL of the participants throughout the entire task. The results support the above discussion, frequent switching between virtual and real content will affect the interruption of the interaction process, thereby reducing the efficiency of VR simulation. Whether in low-interference environment or high-interference environment, the simulation is meaningful when participants focus on the virtual content.

Surprisingly, participants' qualitative feedback was instead focused on the search task. Participant P8 said that "the background environment under VRSS conditions would distract me more easily, thus causing the termination of the search task"; Participant P11 said that "the background environment under VRSS conditions was noisier than expected, which diverted my attention"; Participant P17 said that "wearing a VR display made me feel more tired when searching for words, mainly because it was difficult to focus my eyes. After I found a word, I would accidentally lose it sometimes, and it would take a long time to find the word again." We considered that their cognitive resources could be reallocated, improving their efficiency of use and thus ensuring search time. In addition, the background environment and the virtual screen in VRSS are represented by optical displays, which will weaken the independence between the two. Although this will affect participants' attention, it will increase the sense of immersion and make participants feel more involved in the task, thus offsetting the negative effects for the CWL.In contrast, switching between virtual and real content obviously requires the brain to integrate information structures more, thus increasing CWL.

Regarding the SUS data in the subjective result, there is a significant difference between the VRSS condition and the AR condition only in usability, but not in learnability and total score, the total score of VRSS is lower than the average, which proves that VRSS is still There is a lot of room for improvement. Participant P1 said: "The virtual screen in both the VRSS and AR conditions is not as easy to use as the physical display. The screen is semi-transparent so I can see the background". Participant P3 said: "The virtual screen is not always fixed in the predetermined position and there is a slight jitter and flicker. Participant P20 said, "The clarity of the screen in VR and AR is different from that of physical displays, and the display of text content makes people feel uncomfortable". We conclude that usability is mainly due to inadequate equipment and an incomplete system we have created. The NASA TLX survey was used to measure participants’ perceived workload. The participants only showed significant differences in physical demand, but not in other subjective subscales and total scores. This may be due to more head turns and feeling uncomfortable wearing VR headsets for a long time.

In summary, satisfactory results were not obtained in tasks with frequent switching between virtual and real content, but the overall performance of other tasks was not affected. We believe that the method of using VR to simulate AR is effective and has a guiding significance for designing controlled experiments. We recommend that when designers of AR experiments design control experiments or try to create a scene that does not exist, interaction with real physical objects must be considered.

\subsection{limitations}
There are still many limitations in what we do that need to be addressed. We made the task interface and background environment as similar as possible to using an AR HMD in real life. Concretely, the virtual environment and keyboard setup were designed to resemble current desktop environments. However, 360° videos do not have depth information, and when participants carefully observed the surrounding background environment in VRSS, distortion was still observed. Although participants did not indicate that this had an impact on the task, it still reduced participants' immersion. Therefore, when meeting the requirements of size and price, binocular stereo video and point cloud that can provide better depth of field should still be considered for use. Similarly, real-time images of the mouse and keyboard captured by the camera will also affect the participants' hand-eye coordination. We believe that this sense of disconnection from interacting with physical objects is the main reason why VRSS increases the cognitive load of participants.

Since the background environment needs to be photographed in real time, the AR condition must precede the VRSS condition in the experimental sequence, and the slight learning effect caused by this cannot be avoided. We did not use virtual models as the background environment. Although this simulation method will lead to more work, it should also be considered further. Furthermore, we only studied tasks that involved using a virtual screen to perform office tasks, the scenarios with more interactions with virtual objects are our next research plan.

Secondly, the display devices used in the whole experimental process have some limitations. Despite being the leading commercial AR HMD, HoloLens 2 has unresolved usability issues, including limited FOV, insufficient brightness in strong light, and inaccurate superposition of virtual models on real scenes. Concurrently, Meta Quest 2 also has problems such as insufficient resolution and heavy equipment that puts pressure on the head, and the EEG measurement headband will also affect the comfort of the participants.

Our results may not apply to everyone because we had a small sample size (20 participants) and most of our participants were male college students. This limits the generalizability of the results, as the results may be different in other groups with different gender or age shares. The NASA TLX questionnaire and objective brain frequencies monitoring cannot fully assess CWL, and other analytical methods of the central and peripheral nervous systems (e.g., functional connectivity, heart rate) should also be further considered.

\section{Conclusion}
This paper evaluated the effectiveness of VR simulating AR. We developed a virtual reality simulation system (VRSS) for office tasks using virtual screens, which can simulate the background environment, virtual screen, keyboard, and mouse in the AR scene. To evaluate the usability of the solution, scenes for multiple office tasks using virtual screens were designed, and a user study was conducted with 20 subjects. Participants were required to complete the tasks in AR and then complete them again in VRSS, and their EEG data related to CWL were also recorded.

The results show that frequent switching between virtual and real has an impact on the effect of VRSS. Tasks with less switching can be effectively simulated in both low-interference environment and high-interference environment. Although the feedback expressed by the participants in some tasks was significantly different, it had no significant effect on their CWL. Therefore, we suggest using VR to simulate AR background environments and tasks with less switching between virtual and reality, but avoid frequent visual confirmation to help design control experiments.

In future work, we aim to integrate advances in both software and hardware to develop more realistic and immersive VR methods for simulating AR environments. Our objective is to extend these methods to tasks involving more complex interactions with virtual and physical objects. To achieve this, we plan to design additional experiments, increase the number of participants, and conduct more in-depth analyses of brain activity. These efforts will enable the application of VR simulations of AR to more complex scenes, thereby enhancing the utility and effectiveness of VR in various domains.

\section*{Declaration of competing interest}
The authors declare that they have no known competing financial interests or personal relationships that could have appeared to influence the work reported in this paper. 

\section*{Acknowledgments}
We extend our gratitude to all the volunteers who took part in this experiment. Additionally, we would like to express our appreciation to the anonymous reviewers for their valuable and insightful feedback.

\section*{Funding}
This work was supported by the National Key R\&D Program of China (Grant No. 2023YFB3307903) and the National Natural Science Foundation of China (Grant No. 52275513).

\bibliographystyle{IEEEtran}
\small\bibliography{ref}

\vfill

\end{document}